\documentclass[twoside,12pt]{article}
\usepackage{graphicx}
\usepackage{amssymb}
\def\R{{\mathbb R}}

\def\N{{\mathbb N}}

\def\Z{{\mathbb Z}}
\def\Lat{{\mathbb L}}
\def\supp{{\mbox{\rm supp}\, }}
\def\C{{\mathbb C}}

\def\kasten{$~~\mbox{\hfil\vrule height6pt width5pt depth-1pt}$ }
\newtheorem{theorem}{Theorem}[section]
\newtheorem{assumption}[theorem]{Assumption}
\newtheorem{proposition}[theorem]{Proposition}

\begin{document}
\pagestyle{myheadings} \markboth{H. Gottschalk}{Weakly attractive interaction} \thispagestyle{empty}
\begin{center}
 {\Large \bf Particle systems with weakly attractive interaction}

\

\noindent {\sc Hanno Gottschalk}\\
\vspace{.25cm}

{\small Institut f\"ur angewandte Mathematik, \\ Rheinische
Friedrich-Wilhelms-Universit\"at Bonn,\\ Wegelerstr. 6, D-53115
Bonn, Germany\\ gottscha@wiener.iam.uni-bonn.de}

September 14th 2004
\end{center}

{\noindent \small {\bf Abstract.} Systems of classical continuous particles in the grand canonical ensemble interacting through
purely attractive, yet stable, interactions are defined. By a lattice approximation, FKG ferromagnetic inequalities 
are shown to hold for such particle systems. Using these inequalities, a construction of the infinite volume measures is given by
a monotonicity and upper bound argument. Invariance under Euclidean transformations is proven for the infinite volume measures. The construction 
works for arbitrary activity and temperature and for integrable long range interactions. Also, inhomogeneous systems of particles with different "charge" 
can be treated.  
}

\

{\small \noindent {\bf Keywords}: {\it Weakly attractive interaction, FKG inequality, thermodynamic limit for the ensemble.}
 }
 \section{Introduction}
In the thermodynamics of continuous, classical particles interaction is usually defined through pair potentials. This however does not allow to model
purely attractive interaction, as some kind of repulsive core has to be included into the pair potential in order to render it thermodynamically stable. Similar 
considerations hold for potentials including $n$-point interactions with some upper bound on $n$. 

At the same time, the repulsive core up to now is an obstacle to gain better control over Gibbs measures outside the low-density --- high-temperature (LD-HT) regime. For this regime
Gibbs measures have been constructed rigorously by various kinds of cluster expansions \cite{Ru}, see also \cite{AGY2} for a LD-HT cluster expansion of a class of non-pair potentials. With the exception of \cite{LMP1,LMP2} using Pigorov--Sinai cluster expansion for
low temperatures, the results on Gibbs measures outside of this regime mostly use Banach-Alaoglu like compactness
arguments in combination with tightness to establish existence, see e.g. \cite{Do,GH,Ru2}. Both methods however do not give complete control over the number of limit measures even for a specified set of boundary conditions. 

New applications of random measures and point processes in "soft matter physics", sociology and biology have furthermore increased the need for models that are not based on the rigorous assumption
that the interaction is defined through a pair potential with repulsive core, cf. e.g. \cite{SKM,To}.

Constructions of Gibbs measures for arbitrary parameters have been obtained using GKS ferromagnetic inequalities for (discrete) spin systems or (continuous) two-dimensional
Euclidean quantum field theories \cite{Si}. It is the aim of this paper, which is based on \cite{AGY1,AGY2}, to apply such Euclidean quantum field theory methods in connection with FKG inequalities \cite{FKG} to continuous particle systems. 
FKG inequalities for continuous particle systems have been considered before, see \cite{GK} for a general discussion with applications to Widom-Rowlinson and continuous random cluster type models. Even though in that reference there
is no discussion of the thermodynamic limit, a number of estimates used for the construction of this limit in Section 4 below can, in the special case of only one particle type, also be formulated in terms of the stochastic ordering criterion, which is the main
result of \cite{GK}, see \cite{Go}. In this work, a different method which also extends to more than one type of particles and is based on comparison through interpolation, is used.   

The paper is organized as follows: In Section 2 we define a new  class of purely attractive, yet stable, potentials for systems of continuous, classical particles in the grand canonical ensemble. These are Kac-like potentials given by concave, 
and linearly bounded energy densities in the static field generated by charged particles. Such potentials model attractive forces between particles of the same type and repulsive forces between particles of different type. They could e.g. be useful to describe the attractive part of 
van der Waals  forces of water and hydrocarbon molecules. The linear bound, which
is necessary for stability reasons, then implies that the difference in energy of e.g. a water molecule completely immersed in hydrocarbon molecules and a water
  molecule completely immersed in water molecules is given by a chemical potential, which is reasonable. Compared with non-stable purely attractive pair potentials the saturation of attractiveness between particles of the same type for the potentials used here
none the less can be seen as a weak attractive behavior, which explains the title of this article. On the physical side, these models for the special case of two particle types with opposite charge (or spin) have some similarity with ferrofluid models \cite{GZ,GG,GTZ}. The latter
models however use repulsive forces between particles of all types to obtain stability and are mathematically rather different from what is being discussed here.      
 
In Section 3 ferromagnetic FKG correlation inequalities are proven through an approximation with lattice gas or, in other words, with  spin systems.  These inequalities are then used to construct the themodynamic limit
(TD limit) first on the level of expectations of monotonically increasing function and then on the level of the measures describing the ensemble, see Section 4. The construction works for arbitrary
temperature and density (activity), in particular no low density -- high temperature  conditions are needed. The infinite volume measures are uniquely determined and turn out to be invariant under Euclidean transformations. The space dimension plays no r\^ole and
also long-range interactions can be treated as long as they are integrable.

\section{Weakly attractive interaction}

Let us begin with the heuristics of weakly attractive interaction. For simplicity, in the heuristic discussion 
we only consider one type of (identical) particles in the Euclidean space $\R^d$ with $d\in\N$ the space dimension. 
We assume that each such particle carries a positive unit charge. A charge $s\in\R$ in the point $y\in\R^d$ in the point $x\in\R^d$ gives rise
to a static field $sG(x-y)$, where $G$ is a symmetric ($G(x)=G(-x)$) non-negative function in $L^1(\R^d,dx)$.  The static field $\phi$ generated by a finite number $n$ of particles is given by
$\phi=G*\eta$ where $\eta=\sum_{j=1}^n\delta_{y_j}$ where $y_j\in\R^d$, $j=1,\ldots,n$ give the positions of the particles and $\delta_y$ stands for the Dirac measure in $y$. Let $v:\R\to\R$ be the energy density of the field $\phi$. 
For technical reasons we assume that $v(0)=0$,  $v\in C^2(\R)$ and $v$ has bounded first derivative, $|v'(\phi)|<b$ $\forall \phi\in\R$ and some constant $b>0$. The potential energy of the particle distribution
$\eta$ is given by integration over the energy density of the associated static field $\phi$ 
\begin{equation}
\label{2.1eqa}
U(\eta)=\int_{\R^d} v(\phi)\, dx~~,~~ \phi=G*\eta\, .
\end{equation} 
Let now $v$ be concave, i.e. $v''\leq 0$, cf. Fig.1, and $y\in\R^d$ be the position of a further particle giving rise to the
field $\Delta\phi(x)=G(x-y)$. For the heuristic discussion we assume that $G$ is monotonically decreasing in the distance. Then, if $y$ is far away from the support of $\eta$, the field $\Phi^{\rm ext.}=G*\eta$
can be  neglected in the "range" of $G$ around $y$. The extra particle in $y$ at a point $q$ in this range of $G$ around $y$ then leads to a change of the energy density by $\Delta e_1=v(\Delta\phi(q))$.
Suppose now the points $y$ and $q$ are translated towards the support of $\eta$ where the "exterior" field $\Phi^{\rm ext.}$ can no longer be neglected. The difference in the energy density at the translated point $q'$ is
then given by $\Delta e_2=v(\Phi^{\rm ext.}(q')+\Delta\phi(q))-v(\Phi^{\rm ext.}(q'))$. As $v$ is concave and $\Phi^{\rm ext.}\geq 0$, we get $\Delta e_1\geq\Delta e_2$, see Fig. 1. Hence, the particle in $y$ gains energy if it approaches the support of $\eta$. 
This explains the attractive character of the potential introduced in (\ref{2.1eqa}) for concave $v$. The potentials of this type are called weakly attractive potentials, because, due to the linear 
asymptotics of $v$, there is a saturation of the attractiveness for the single particle in regions which are homogenuously filled by a very strong "exterior" field $\Phi^{\rm ext.}$, see again Fig. 1.  
\setlength{\unitlength}{1cm}
\begin{figure}[ht]
\begin{picture}(8,12)
\put(2,0){\vector(0,1){8.5}}
\put(0,7){\vector(1,0){11.5}}
\put(.3,8.5){$e=v(\phi)$}
\put(.4,8){energy}
\put(.4,7.5){density}
\put(11.2,7.25){$\phi$}
\bezier{55}(7.2,6.8)(9.2,3.4)(11.2,0)
\thicklines
\bezier{200}(2,7)(2.5,6.99)(3,6.97)
\bezier{200}(3,6.97)(3.5,6.93)(4,6.88)
\bezier{200}(4,6.88)(4.5,6.8)(5,6.7)
\bezier{200}(5,6.7)(5.5,6.5)(6,6.3)
\bezier{200}(6,6.3)(6.5,6)(7,5.6)
\bezier{200}(7,5.6)(7.5,5.1)(8,4.6)
\bezier{200}(8,4.6)(8.5,4)(9,3.3)
\bezier{200}(9,3.3)(9.5,2.6)(10,1.8)
\bezier{200}(10,1.8)(10.5,0.95)(11,.1)
\linethickness{.4mm}
\put(2,7){\vector(1,0){3}}
\put(7,7){\vector(1,0){3}}
\put(3.25,7.2){$\Delta\phi$}
\put(8.25,7.2){$\Delta\phi$}
\thicklines
\bezier{10}(7,7)(7,6.3)(7,5.6)
\bezier{20}(7,5.6)(8.5,5.6)(10,5.6)
\bezier{10}(10,7)(10,6.3)(10,5.6)
\put(2,6.9){\line(0,1){.2}}
\put(7,6.9){\line(0,1){.2}}
\put(6.85,7.2){$\Phi^{\rm ext.}$}
\thinlines
\put(5,7){\vector(0,-1){.3}}
\put(5.3,6.65){$\Delta e_1$}
\put(10.02,5.6){\vector(0,-1){3.7}}
\put(10.3,3.7){$\Delta e_2$}
\end{picture}

\caption{A model for weakly attractive interaction}
\end{figure}
If we allow the particles in $y_1,\ldots,y_n$ to carry charges $s_1,\ldots,s_n\in R$, hence $\eta=\sum_{j=1}^ns_j\,\delta_{y_j}$, $\phi=G*\eta$. It is then clear from the above discussion that weakly attractive potentials, as defined
by (\ref{2.1eqa}) with $v$ concave, lead to attractive interaction
between particles of the same charge and to repulsive forces between particles of different charges. 

In a thermodynamic ensemble of particles with weakly attractive interaction, groupings of particles with the same sign of the charge
are energetically preferred w.r.t. mixtures of particles with different signs of the charge. This leads to positive correlation between the event of finding a particle of given charge in a certain region and the
event of finding another particle nearby with the same sign of the charge. In technical terms, this should be expressed by a "ferromagnetic" correlation inequality.
On the other hand, the saturation of attractiveness in regions of strong fields of either sign implies that the number of particles in a given volume can always be controlled (also in the TD limit) by the number of particles
of a free gas with higher activity. These two ingredients -- ferromagnetic correlation inequalities, implying monotonicity of the expectation values for monotonically increasing function in the TD limit, and upper bounds of the particle number due to comparison
with a free gas of higher activity -- is essentially all what is needed to copy Nelson's classical proof of the TD limit of $P(\phi)_2$-theories \cite{Si}.

 Having clarified the heuristic basis, we now carry on with the technical formulation following \cite{AGY1,AGY2}. Let $z>0$ be the activity and let $r$ be a probability measure on $\R$ with compact support, $\supp r\subseteq [-C,C]$
 for some $C>0$, such that $r\{0\}=0$. We define a purely Poisson L\'evy characteristic 
 $\psi(t)=z\int_\R[e^{ist}-1]\,dr(s)$, $t\in\R$, and we set for $f\in{\cal S}(\R^d)$, where ${\cal S}(\R^d)$ is the space of real Schwartz test functions on $\R^d$, 
 \begin{equation}
 \label{2.2eqa}
 {\cal C}_0(f)=e^{\int_{\R^d}\psi(f)\, dx} \, .
 \end{equation} 
 Then, ${\cal C}_0:{\cal S}(\R^d)\to \C$ is a positive definite, normalized functional which is also continuous w.r.t. the nuclear Schwarz topology on ${\cal S}(\R^d)$. Let ${\cal S}'(\R^d)$ be the space of tempered distributions over $\R^d$ (i.e. the topological dual space of 
 ${\cal S}(\R^d)$) and let ${\cal B}={\cal B}({\cal S}'(\R^d))$ be the Borel sigma algebra (the sigma algebra generated by open sets) over ${\cal S}'(\R^d)$.
 By Minlos' theorem \cite{AGW1,M}, there exists a unique probability measure $\mu_0$ on the measurable space $({\cal S}'(\R^d),{\cal B})$ such that  
 \begin{equation}
 \label{2.3eqa}
 {\cal C}_0(f)=\int_{{\cal S}'(\R^d)}e^{i\langle\eta,f\rangle}\, d\mu_0(\eta)~~\forall f\in{\cal S}(\R^d)\, .
 \end{equation}
 
The measure $\mu_0$ has a natural interpretation as the Gibbs measure of a non-interacting system of classical, continuous particles in the grand canonical ensemble where each particle carries a random charge
distributed according to $r$. In fact, for $\Lambda\subseteq \R^d$ with finite Lebesgue volume, $|\Lambda|<\infty$, let $N_\Lambda^z$ be a Poisson random variable (number of particles in $\Lambda$) with intensity $z|\Lambda|$ and $\{Y_j\}_{j\in\N}$,
$\{S_j\}_{j\in\N}$  two families of i.i.d. random variables (also mutually independent and independent from $N_\Lambda^z$) where $Y_j$ is uniformly distributed in $\Lambda$ (position of the $j$th particle in $\Lambda$)
and $S_j$ (charge of the $j$th particle in $\Lambda$) has distribution according to $r$, then
the restriction of the coordinate process $\langle.,f\rangle$ to $\Lambda$, which maps test functions with support in $\Lambda$ to random variables on $({\cal S}'(\R^d),{\cal B},\mu_0)$ is equivalent in law with the random process ${\cal S}(\Lambda)\ni
f\to\sum_{j=1}^{N_\Lambda^z}S_j\,f(Y_j)=\langle \eta_{0,\Lambda},f\rangle$ where $\eta_{0,\Lambda}=\sum_{j=1}^{N_\Lambda^z}S_j\,\delta_{Y_j}$. 

For the measure $\mu_0$ this together with a short estimate \cite{AGY2} implies that $\mu_0(\Gamma)=1$ where $\Gamma=\Gamma^C\subseteq{\cal S}'(\R^d)$ is the set of all signed measures with only finitely many support points in any compact subset of $\R^d$
s.t. for $\eta\in\Gamma$ the following holds: $|\eta\{y\}|\leq C$ $\forall y\in\R^d$ and $\int_{\R^d}(1+|x|^2)^{-(d+\epsilon)/2}d|\eta|(x)<\infty$ $\forall \epsilon>0$. For $\eta=\sum_{y\in{\rm supp}\eta}s_y\delta_y$, $|\eta|=\sum_{y\in{\rm supp}\eta}|s_y|\delta_y$.
$\Gamma$ is in fact a measurable set in $({\cal S}'(\R^d),{\cal B})$, cf. \cite{AGY2}. In the probability space $({\cal S}'(\R^d),{\cal B},\mu_0)$ one can replace ${\cal S}'(\R^d)$ with $\Gamma$ and ${\cal B}$ with ${\cal B}^\Gamma={\cal B}\cap \Gamma$, the
trace sigma algebra of ${\cal B}$ on $\Gamma$. 

These support properties of $\mu_0$ imply the following: Let the kernel function $G$ be as above.  One can show that for $\eta\in \Gamma$, 
$G*\eta\in L^1_{\rm loc}(\R^d,dx)$ \cite{AGY2}. For a infra-red cut-off function with compact support $g\in C_0(\R^d)$, $g\geq 0$, one can thus define
the following interaction
\begin{equation}
\label{2.4eqa}
U_g(\eta)=\int_{\R^d}v(\phi)\, g dx\, ,~~\phi=G*\eta.
\end{equation}
Here $v$ fulfills the conditions given above, in particular $v$ is concave. That we defined the infra-red cut-off in this way instead of using finite volume measures $\mu_{0,\Lambda}$ and the interaction
(\ref{2.1eqa}) is only a technicality, which is going to allow us to keep as close to Nelson's proof of the TD limit for $P(\phi)_2$ as possible.

From \cite{AGY2} we now get:

\begin{proposition}
\label{2.1prop}
(i) $U_g:\Gamma\to\R$ is ${\cal B}^\Gamma$ measurable;

\noindent (ii) $U_g\in L^{p}(\Gamma,\mu_0)$ $\forall\, p\geq 1$;

\noindent (iii) $e^{-U_g}\in L^{p}(\Gamma,\mu_0)$ $\forall\, p\geq 1$.  
\end{proposition}   
\noindent {\bf Sketch of the proof:} The crucial point in the proof of Proposition \ref{2.1prop} is the very simple estimate
\begin{equation}
\label{2.5eqa}
|U_g(\eta)|\leq b\int_{\R^d}G*g\, d|\eta|
\end{equation}  
and the observation that 
\begin{equation}
\label{2.6eqa}
\int_\Gamma e^{b\langle|\eta|,G*g\rangle}d\mu_0(\eta)<\infty
\end{equation}
where the latter inequality follows from calculating the Laplace transform of the
measure $\mu_0\circ \varphi^{-1}_+$ with $\varphi_+(\eta)=|\eta|$, cf. (\ref{4.6eqa}) below. For the details see \cite{AGY2}. \kasten

By Proposition \ref{2.1prop} (iii) the following interacting Gibbs measures $\mu_g$ with infra-red cut-off $g$ are well-defined on $(\Gamma,{\cal B}^\Gamma)$:
\begin{equation}
\label{2.7eqa}
d\mu_g(\eta)={e^{-U_g(\eta)}\over\int_\Gamma e^{-U_g}\, d\mu_0}\, d\mu_0(\eta)\,.
\end{equation} 
The aim of the paper is now to remove the infra-red cut-off $g$, i.e. for inverse temperature $\beta>0$ we want to investigate
the limit $g\nearrow\beta$ for $g\in C_0(\R^d)$ for the measures $\mu_g$ and show their convergence to a limit measure $\mu_\beta$. Before we can resume with
this task in Section 4, ferromagnetic FKG inequalities have to be established through lattice approximation, which is done in the following section.   
 
\section{Lattice approximation and FKG inequality}

First let us introduce the FKG (for Fortuin, Kasteleyn and Ginibre) inequalities and state the main result of this section:

Let ${\cal O}(\Gamma)$ be a specific set of functions -- also called observables -- i.e. measurable functions $F:\Gamma\to \R$. Namely, for $F\in{\cal O}(\Gamma)$ there exists $n\in\N$, a continuous function $H:\R^{n}\to\R$ exponentially bounded and fast falling
functions $h_1,\ldots,h_n\in C_{\rm f.f.}(\R^d)$ such that $F(\eta)=H(\langle\eta,h_1\rangle,\ldots \langle\eta,h_n\rangle)$. The fast falling functions are defined as $C_{\rm f.f}(\R^d)=\{h\in C(\R^d):\sup_{x\in\R^d}|(1+|x|^2)^Nh(x)|<\infty$ $\forall N\in\N\}$ and exponentially bounded means that 
there exist $\kappa,K>0$ s.t. $H(x_1,\ldots,x_n)\leq Ke^{\kappa (|x_1|+\cdots+|x_n|)}$ $\forall x\in\R^n$. 
From the definition of $\Gamma$ it is clear that $\langle\eta,h\rangle$ is well-defined for $\eta\in\Gamma, h\in C_{\rm f.f}(\R^d)$. By the exponential boundedness of the functions $H$, $|F(\eta)|\leq Ke^{\kappa\langle|\eta|,h\rangle}$, $h=|h_1|+\cdots+|h_n|\in C_{\rm f.f.}(\R^d)$ and by an estimate like (\ref{2.6eqa})
it follows that ${\cal O}(\Gamma)\subseteq L^{p}(\Gamma,\mu_0)$ $\forall p\geq 1$. Proposition \ref{2.1prop} (iii) also implies
${\cal O}(\Gamma)\subseteq L^p(\Gamma,\mu_g)$, $p\geq 1$.

An observable $F\in{\cal O}(\Gamma)$ is called monotonically increasing if $F$ can be represented by a function $H:\R^{n}\to\R$ and $h_1,\ldots, h_n\in C_{\rm f.f.}(\R^d)$ such that
$h_l\geq0$, $l=1,\ldots,n$, and $H(x_1,\ldots,x_n)$ is monotonically increasing in each argument $x_1,\ldots,x_n$. By $\hat {\cal O}(\Gamma)$ we denote the collection of all monotonically increasing functions $F\in{\cal O}(\Gamma)$. By definition,
 a probability measure $\mu$ on $(\Gamma,{\cal B}^\Gamma)$ fulfills the FKG inequality, if
 \begin{equation}
 \label{3.1eqa}
 \int_\Gamma F_1F_2 \, d\mu\geq \int_\Gamma F_1 \, d\mu\int_\Gamma F_2 \, d\mu~~\forall\, F_1,F_2\in\hat {\cal O}(\Gamma).
 \end{equation}
\begin{theorem}
\label{3.1theo}
The FKG inequality (\ref{3.1eqa}) holds for $\mu_g$ defined as in Section 2.
\end{theorem}

The proof of the FKG inequality for the gas of continuous particles in the GCE with weakly attractive interaction requires various steps of approximation preserving the FKG inequality. How the approximations
are done in detail to a large extend is a matter of convenience. Here we do the following steps: 
\begin{itemize}
\item Approximate the continuous system with kernel $G\in L^1(\R^d,dx)$ as specified in the previous section by a continuous system with finite range interactions $G\in C_0(\R^d)$, $G\geq 0$ (Proposition \ref{3.1prop});
\item Approximate the continuous system with the regular and finite-range kernel $G$ by infinite lattice systems (Proposition \ref{3.2prop});
\item Approximate infinite lattice systems by finite lattice systems (Proposition \ref{3.3prop});
\item Verify the FKG inequality for the finite lattice system (Proposition \ref{3.4prop}).
\end{itemize}
The technical details of the different approximation always follow the same \linebreak scheme, namely a proof of convergence in $L^p(\Gamma,\mu_0)$ through pointwise convergence and a $L^p(\Gamma,\mu_0)$-upper bound
through an estimate as in the proof of Proposition \ref{2.1prop}. To avoid repetitions, we give the detailed proof for the first steps only and then indicate where this scheme has to be modified for the
remaining steps.    
 
Let $G$ be as described in the previous section and let $G_n\in C_0(\R^d)$, $G_n\geq 0$, and
$G_n\to G$ in $L^1(\R^d,dx)$ as $n\to\infty$ such that $|G_n|\leq \tilde G$ for a $L^1(\R^d,dx)$ majorant $\tilde G$.
 Let $U_{g,n}$ be defined as in (\ref{2.4eqa}) with $G$ replaced by $G_n$. Then
\begin{proposition} 
\label{3.1prop}

\noindent  (i) $U_{g,n}\to U_g$ in $L^p(\Gamma,\mu_0)$ as $n\to\infty$ for $p\geq 1$;

\noindent (ii) $e^{-U_{g,n}}\to e^{-U_g}$ in $L^p(\Gamma,\mu_0)$ as $n\to\infty$ for $p\geq 1$.
\end{proposition}
\noindent {\bf Proof.} We notify that
\begin{equation}
\label{3.2eqa}
\left| U_g(\eta)-U_{g,n}(\eta)\right|\leq b\int_{\R^d} |G-G_n|*g\,d|\eta|\, , ~~\eta\in\Gamma
\end{equation}
$G_n\to G$ in $L^1(\R^d,dx)$ and $g\in C_0(\R^d)$ imply $|G-G_n|*g\to 0$ pointwise in $\R^d$ as $n\to\infty$.
By (\ref{2.6eqa}), $\int_{\R^d} \tilde G*g\, d|\eta|<\infty$ for $\mu_0$ almost all $\eta\in\Gamma$. Thus, the right hand side
of (\ref{3.2eqa}) converges to zero by the theorem of dominated convergence applied on $L^1(\R^d,d|\eta|)$ for $\mu_0$ almost all $\eta\in\Gamma$ as $n\to\infty$. 

This implies that $U_{g,n}\to U_g$ and $e^{-U_{g,n}}\to e^{-U_g}$ $\mu_0$ a.s. as $n\to\infty$. The assertion (ii) now follows from the theorem of dominated convergence in $L^{p}(\Gamma,\mu_0)$. 
In fact, $e^{-U_{g,n}(\eta)}\leq e^{b\langle|\eta|,\tilde G*g\rangle}$ uniformly in $n$ and
the right hand side of this estimate is in $L^{p}(\Gamma,\mu_0)$ for $p\geq1$, cf. (\ref{2.6eqa}) where $b$ can be chosen arbitrarily. 
The assertion (i) follows from the fact that the upper bound in (ii) can always be taken as an upper bound for $U_{g,n}$.
\kasten

In the remainder of this section we thus assume $G\in C_0(\R^d)$, $G\geq 0$, if not defied differently. The next step is the lattice approximation:
 
Let $\lambda>0$ be the lattice spacing for the lattice $\lambda\Z^d\subseteq\R^d$. We consider each point of this lattice as center of a box $\Lambda_j^\lambda$, $j\in\lambda\Z$, which has side length 
$\lambda$. The boundaries of the boxes are attached to either of the boxes having these boundaries in common, such that the $\Lambda^\lambda_j$ give a disjoint partition of $\R^d$.
Given $\eta \in \Gamma$, we define $\eta_j^\lambda=\eta(\Lambda_j^\lambda)$. Define $\eta^\lambda=\sum_{j\in \lambda\Z^d}\lambda^{-d}\eta_j^\lambda\, 1_{\Lambda_j^\lambda}$ with $1_A$ the indicator
function of the set $A\subseteq \R^d$.The lattice interaction $U_g^\lambda:\Gamma\to\R$ is defined through
\begin{equation}
\label{3.3eqa}
U_g^{\lambda}(\eta)=\int_{\R^d} v(\phi^\lambda)\, gdx\, ,~~\phi^\lambda=G*\eta^\lambda\, . 
\end{equation}  
By a simple adaptation of the proof of Proposition \ref{2.1prop} to the discretized case one immediately gets that the statements (i)-(iii) of Proposition \ref{2.1prop} remain true if $U_g$
is replaced by $U_g^\lambda$. Furthermore, for $F\in{\cal O}(\Gamma)$ defined through $F(\eta)=H(\langle\eta, h_1\rangle,\ldots,\langle\eta, h_n\rangle)$ as described above, we obtain $F^\lambda$
by replacing $\eta$ by $\eta^\lambda$. That this procedure might depend on $H$ and $h_1,\ldots,h_n$ and not only on $F$, is not an obstacle: For any $F\in{\cal O}(\Gamma)$ we {\em a priori} fix one
such representation.  We get the following convergence of the lattice potential and the lattice observables:

\begin{proposition}
\label{3.2prop}

\noindent (i) $U_g^\lambda\to U_g$ in $L^p(\Gamma,\mu_0)$ as $\lambda \searrow 0$ for $p\geq 1$;

\noindent (ii) $e^{-U_g^\lambda}\to e^{-U_g}$ as $\lambda \searrow 0$ in $L^p(\Gamma,\mu_0)$ for $p\geq 1$;

\noindent (iii) $F^\lambda\to F$ as $\lambda \searrow 0$ in $L^p(\Gamma,\mu_0)$ for $p\geq 1$.
\end{proposition}
\noindent {\bf Proof.} Let $h\in C_{\rm f.f.}(\R^d)$ and $\eta\in\Gamma$,
 $\eta=\sum_{y\in{\rm supp} \eta}s_y\,\delta_y$. Then,
\begin{eqnarray}
\label{3.4eqa}
\left\langle\eta^\lambda,h\right\rangle&=&\sum_{j\in\lambda\Z^d}\lambda^{-d}\eta_j^\lambda\int_{\Lambda_j^\lambda}h(x)\, dx \nonumber\\
&=& \sum_{j\in\lambda\Z^d}\sum_{y\in{\rm supp} \eta\cap \Lambda_j^\lambda}s_y\, \lambda^{-d}\int_{\Lambda_j^\lambda}h(x)\, dx\nonumber\\
&=& \sum_{y\in{\rm supp} \eta} s_y \, \lambda^{-d}\int_{\Lambda^\lambda _{j^\lambda(y)}}h(x)\, dx \to\langle\eta,h\rangle\, \mbox{ as } \lambda\searrow 0.
\end{eqnarray}
Here $j^\lambda(y)$ for $y\in\supp \eta$ is defined to be $j\in\lambda\Z$ s.t. $y\in\Lambda_j^\lambda$. The sums all converge absolutely by definition of $\Gamma$ and $C_{\rm f.f.}(\R^d)$. The
convergence as $\lambda\searrow 0$ follows from $\lambda^{-d}\int_{\Lambda_{j^\lambda(y)}^\lambda}h(x) \, dx\to h(y)$ and a dominated convergence argument using the rapid decay of $h$ and $|s_y|\leq C$. 

We first prove (i) and (ii). Note that $\phi(x)=G*\eta(x)=\langle\eta,G_x\rangle$, $G_x(y)=G(x-y)$, and $G_x\in C_{\rm f.f.}(\R^d)$ and the same holds for $\phi^\lambda(x)$ with $\eta$
replaced by $\eta^\lambda$. By (\ref{3.4eqa}), 
$\phi^\lambda(x)\to\phi(x)$ $\forall x\in\R^d$. Consequently, $v(\phi^\lambda)\to v(\phi)$ in $L^1(\R^d,g\, dx)$ as 
\begin{equation}
\label{3.5eqa}
|v(\phi^\lambda(x))|\leq bC \|G\|_\infty \sharp (\Lambda(g,G,\lambda)\cap \supp\eta)~~~\forall x\in\supp g
\end{equation}
with $b$ the linear bound of $v$, $C$ from the definition of $\Gamma$ and $\Lambda(g,G,\lambda)=\{y\in\R^d:\inf_{x\in({\rm supp} g+{\rm supp} G)}|y-x|\leq\lambda\}$ and the term on the right hand side is monotonically falling
in $\lambda$. $\sharp A$ denotes the number of elements in a set $A$. Thus, $U_g^\lambda\to U_g$ pointwisely in $\Gamma$. A $L^p(\Gamma,\mu_0)$--upper bound for $U_g^\lambda$, $\lambda \leq 1$, and $e^{-U_g^\lambda}$ is $e^{\alpha N_{\Lambda(G,g,1)}^z}$
with $\alpha=bC\|G\|_\infty\|g\|_{L^1(\R^d,dx)}$, $N^z_\Lambda(\eta)=\sharp (\Lambda\cap\supp\eta)$ for $\Lambda\subseteq\R^d$, cf. (\ref{3.5eqa}).    

Let us now consider (iii). $F^\lambda\to F$ pointwisely on $\Gamma$ is a consequence of (\ref{3.4eqa}) and the continuity of $H:\R^{dn}\to \R$. The $L^p(\Gamma,\mu_0)$--upper bound exists because of
the exponential boundedness of $H$ and the estimate $|\langle\eta^\lambda,h\rangle|\leq B\sup_{x\in\R^d}|h(x)(1+|x|^2)^{N/2}|\langle|\eta|,(1+|.|^2)^{-N/2}\rangle$, for $N>d$, $\lambda<1$ and $B>0$ large enough, where the 
exponential of the r.h.s. is in $L^p(\Gamma,\mu_0)$ for $p\geq 1$, cf. (\ref{2.6eqa}) and \cite{AGY2}.  
\kasten

The next step is the approximation of $\eta^\lambda$ by $\eta^\lambda_\Lambda=\sum_{j\in\lambda\Z^d\cap\Lambda}\lambda^{-d}\eta_j^\lambda 1_{\Lambda_j^\lambda}$ as $\Lambda\subseteq\R^d$ compact approaches $\R^d$. The notation
$\Lambda\nearrow\R^d$ is used for this limit; the precise meaning of $\lim_{\Lambda\nearrow\R^d}$ is that for any sequence $\Lambda_n\subseteq \R^d$ compact such that $\Lambda_n\subseteq \Lambda_m$ for $n\leq m$ and $\cup_{n\in\N}\Lambda_n=\R^d$
the limit exists and all such limits coincide. $U_{g,\Lambda}^\lambda(\eta)=\int_{\R^d}v(\phi_\Lambda^\lambda)\, gdx$, $\phi^{\lambda}_\Lambda=G*\eta_\Lambda^\lambda$, and $F^\lambda_\Lambda$ is defined through a representation $H:\R^{dn}\to\R$, 
$h_1,\ldots,h_n\in C_{\rm f.f.}(\R^d)$ by replacing $\langle\eta,h_l\rangle$ with $\langle\eta_\Lambda^\lambda,h\rangle$, $l=1,\ldots,n$.
\begin{proposition}
\label{3.3prop}

\noindent (i) $U_{g,\Lambda}^\lambda\to U_g^\lambda$ in $L^p(\Gamma,\mu_0)$ as $\Lambda \nearrow \R^d$ for $p\geq 1$;

\noindent (ii) $e^{-U_{g,\Lambda}^\lambda}\to e^{-U_g^\lambda}$ as $\Lambda \nearrow \R^d$ in $L^p(\Gamma,\mu_0)$ for $p\geq 1$;

\noindent (iii) $F^\lambda_\Lambda\to F^\lambda$ as $\Lambda \nearrow \R^d$ in $L^p(\Gamma,\mu_0)$ for $p\geq 1$.
 
\end{proposition}
{\bf Proof.} Firstly we notify that the upper $L^p(\Gamma,\mu_0)$--bounds given in Proposition \ref{3.2prop} for $U_g^\lambda$ $e^{-U_g^\lambda}$ and $F^\lambda$
are upper $L^p(\Gamma,\mu_0)$--bounds also for $U_{g,\Lambda}^\lambda$, $e^{-U_{g,\Lambda}^\lambda}$ and $F_\Lambda^\lambda$ uniformly in $\Lambda\subseteq \R^d$ 
(strictly speaking this holds for $\lambda\leq 1$ only  but related bounds
can be found for $\lambda>1$). It remains to prove pointwise convergence.

To prove (i) and (ii) we use the finite range of $G$ and we obtain $\phi_\Lambda^\lambda(x)\to \phi^\lambda(x)$ as $\Lambda\nearrow\R^d$ where the limit for fixed $x\in\R^d$
is obtained increasing $\Lambda$ in the limit $\Lambda\nearrow\R^d$ finitely many times. Thus, $v(\phi_\Lambda^\lambda)\to v(\phi^\lambda)$ on $\R^d$ pointwisely. With the same $L^1(\R^d,gdx)$--upper bounds as 
in Proposition \ref{3.2prop} this implies $U_{g,\Lambda}^\lambda\to U_g^\lambda$ on all $\Gamma$.

It remains to prove (iii). The convergence $\langle\eta_\Lambda^\lambda,h\rangle\to\langle\eta^\lambda,h\rangle$ for $\eta\in\Gamma$, $h\in C_{\rm f.f.}(\R^d)$ as $\Lambda\nearrow\R^d$, is a simple
consequence of the definition of these spaces which permits to get the statement by the theorem dominated convergence in $L^1(\R^d,dx)$. This implies the pointwise convergence $F^\lambda_\Lambda\to F^\lambda$ on $\Gamma$
as $\Lambda\nearrow\R^d$. \kasten

Until now we have reduced the system to a finite lattice system where the state space can be identified with $\R^m$ with $m=m(\Lambda,\lambda)=\sharp \Lat$, $\Lat=\Lat(\Lambda,\lambda)=\Lambda\cap\lambda\Z^d$. Before proceeding to the FKG inequality, we want to make this
more explicit. In fact, for $F\in{\cal O}(\Gamma)$ with fixed representation
$F(\eta)=H(\langle\eta,h_1\rangle,\ldots,\langle\eta,h_n\rangle)$. Thus, keeping $\Lambda\subseteq\R^d$ compact and $\lambda>0$ fixed, 
\begin{eqnarray}
\label{3.6eqa}
F_\Lambda^\lambda(\eta)&=&H\left(\sum_{j\in\Lambda\cap\lambda\Z^d}\lambda^{-d}\eta_j^\lambda\langle 1_{\Lambda_j^\lambda},h_1\rangle,\ldots,\sum_{j\in\Lambda\cap\lambda\Z^d}\lambda^{-d}\eta_j^\lambda\langle 1_{\Lambda_j^\lambda},h_n\rangle\right)\nonumber\\
&=&\tilde H(\eta_j^\lambda|j\in\Lat).
\end{eqnarray}
We notify that for $F\in\hat {\cal O}(\Gamma)$ one can choose $\tilde H\in\hat{\cal O}(\R^m)$ where the latter space is defined as the space of
exponentially bounded, continuous functions from $\R^m$ to $\R$ monotonically increasing in each argument $x_j\in\R$, $j=1,\ldots,m$. For differentiable functions $H$ this follows by
differentiation w.r.t. $\eta_j^\lambda$ using the chain rule and $\lambda^{-d}\langle1_{\Lambda_j^\lambda},h_l\rangle\geq 0$
if $h_l\geq 0$. For only continuous $H$ this follows by approximation with differentiable functions.   	   

We note that the family of random variables $\{\eta_j^\lambda\}_{j\in\Lat}$ is i.i.d distributed. Let $\rho$ be this distribution on $\R$, i.e. $\rho$ is the unique
probability measure on $(\R,{\cal B}(\R))$ with Fourier transform  
$\int_{\R}e^{ist}\, d\rho(s)=\int_{\Gamma}e^{it\langle\eta,1_{\Lambda_j^\lambda}\rangle}\, d\mu_0(\eta)=e^{\lambda^{d}\psi(t)}$,
cf. (\ref{2.2eqa}) and note that $\psi(0)=0$. We can now define the lattice measure $\mu_g^\Lat$ on $(\R^m,{\cal B}(\R^m))$ through
\begin{equation}
\label{3.7eqa}
d\mu_g^\Lat(\times_{j\in\Lat}\eta_j)={e^{-U^\Lat_g(\eta_j|j\in\Lat)}\over \int_{\R^m}e^{-U^\Lat_g(\eta_j|j\in\Lat)}d\rho^{\otimes m}(\times_{j\in\Lat} \eta_j)}d\rho^{\otimes m}(\times_{j\in\Lat} \eta_j)\, ,
\end{equation}
with $U^\Lat_g(\eta_j)=U_{g,\Lambda}^\lambda(\eta)$ for $\eta_j=\eta^\lambda_j$, $j\in\Lat$, i.e. 
$U^\Lat_g(\eta_j|j\in\Lat)=\int_{\R^d}v(\phi_\Lat)\, gdx$, $\phi_\Lat=\sum_{j\in\Lat}\lambda^{-d}\eta_j\, G*1_{\Lambda_j^\lambda}$.

These definitions imply
\begin{equation}
\label{3.8eqa}
\int_{\R^m}\tilde H\, d\mu_g^\Lat={\int_\Gamma F_\Lambda^\lambda e^{-U_{g,\Lambda}^\lambda}d\mu_0\over \int_\Gamma e^{-U_{g,\Lambda}^\lambda}d\mu_0}
\end{equation}
for any $\tilde H:\R^m\to\R$ such that (\ref{3.6eqa}) holds. For $\mu_g^\Lat$ we get the following result:

\begin{proposition}
\label{3.4prop}
$\mu_g^\Lat$ fulfills the FKG inequality on $\R^m$, i.e. 
\begin{equation}
\label{3.9eqa}
\int_{\R^m}\tilde H_1\tilde H_2\, d\mu_g^\Lat\geq \int_{\R^m}\tilde H_1\, d\mu_g^\Lat\int_{\R^m}\tilde H_2 \, d\mu_g^\Lat~~\forall~\tilde H_1,\tilde H_2\in\hat {\cal O}(\R^m).
\end{equation}
\end{proposition}
Clearly, (\ref{3.9eqa}) completes the proof of Theorem \ref{3.1theo}: For $F_l\in\hat {\cal O}(\Gamma)$, $l=1,2$, one can choose the
representing $\tilde H_l$ from $\hat{\cal O}(\R^m)$ and from Prop. \ref{3.4prop} and (\ref{3.8eqa}) one obtains the correlation inequality for the measure defined on the right hand side 
of that equation for observables $F_{l,\Lambda}^\lambda$, $l=1,2$ and arbitrary $\lambda>0$, $\Lambda\subseteq \R^d$ compact. Using Propositions \ref{3.2prop}--\ref{3.3prop} (ii), (iii) and \ref{3.1prop} (ii) one gets the convergence of both sides of the so-obtained inequality
to the respective sides of (\ref{3.1eqa}) with $\mu=\mu_g$ if we let first $\Lambda\nearrow\R^d$ then $\lambda \searrow 0$ and finally carry out the short range approximation where in every step integrands in the numerators and denominators converge in $L^p(\Gamma,d\mu_0)$. 
The statements Proposition \ref{3.1prop}--\ref{3.3prop} (i)
are not needed here, but in the following section.

\noindent {\bf Proof of Prop. \ref{3.4prop}.} We first assume that the measure $r$ defining $\psi$ is absolutely continuous w.r.t. to the Lebesgue measure
$ds$ on $\R$ with a $C^2_0(\R)$-density function and $\supp r\cap(0,\infty)\not=\emptyset$, $\supp r\cap (-\infty,0)\not=\emptyset$. By investigation of the Fourier transform of $\rho$ one gets: For $f\in C(\R)$
exponentially bounded 
\begin{equation}
\label{3.10eqa}
\int_{\R}f\, d\rho=e^{-z\lambda^d}\sum_{n=1}^\infty{(z\lambda^d )^n\over n!}\int_\R f\, dr^{*n}
\end{equation}
 where $r^{*n}$ is the $n$-fold convolution of $r$ with itself. 

 This implies that $\rho$ is absolutely continuous w.r.t. the Lebesgue measure, $d\rho(\eta_j)=\varrho(\eta_j)d\eta_j$ with
$\varrho\in C^2(\R)$ and $\varrho>0$ on $\R$, and $\log\varrho$ is a well-defined function in $C^2(\R)$. Consequently, 
\begin{equation}
\label{3.11eqa}
d\mu_g^\Lat(\times_{j\in\Lat}\eta_j)=e^{-W(\eta_j|j\in\Lat)}\otimes_{j\in\Lat} d\eta_j
\end{equation}
with $W(\eta_j|j\in\Lat)=U_g^\Lat(\eta_j|j\in\Lat)-\sum_{j\in\Lat}\log\varrho(\eta_j)-\log\Xi$, $\Xi$ being the normalization constant $\int_{\R^m}e^{-W(\eta_j|j\in\Lat)}\otimes_{j\in\Lat} d\eta_j$. 
 In order to verify the FKG inequality for a 
probability measure of type (\ref{3.11eqa}) with $W:\R^m\to\R$ in $C^2(\R^m)$, it suffices
to verify the logarithmic FKG-criterium $\left. \partial^2W(\eta_j|j\in\Lat)\right/\partial\eta_j\partial\eta_l\leq 0$ on $\R^m$ for $l\not=j$, $l,j\in\Lat$ \cite{Si2} .
In our case 2nd order partial derivatives of $W$ exist as $v\in C^2(\R)$ and $\log\varrho\in C^2(\R^d)$. Carrying out the partial differentiations one gets from the definition of $U_g^\Lat$
\begin{equation}
\label{3.12eqa}
{\partial^2 W(\eta_j|j\in\Lat)\over \partial \eta_j\partial \eta_l}=\lambda^{-2d}\int_{\R^d}v''(\phi_\Lat)\, G*1_{\Lambda_j^\lambda} G*1_{\Lambda_l^\lambda}\, gdx\leq 0
\end{equation}  
for $j,l\in\Lat$, $j\not=l$ and $\phi_\Lat$ defined in terms of the variables $\eta_j$ as above. The crucial assumptions that $v$ is concave, $v''\leq 0$, and $G\geq 0$ enter only here. 

For the general case, where $r$ is not necessarily absolutely continuous w.r.t. $ds$, we use the following approximation preserving the FKG inequality: Let $dr_\epsilon
=[(1-\epsilon)(\chi_\epsilon*r)+\epsilon(\chi_\epsilon*r^\theta)]\, ds$ where $r^\theta(A)=r(-A)$, $A\in{\cal B}(\R)$. $\chi\in C_0^2(\R)$ is a symmetric mollifyer, i.e. $\chi\geq 0$, $\chi(-s)=\chi(s)$ and $\int_\R\chi\,ds=1$, $\chi_\epsilon(s)=\chi(\epsilon s)/\epsilon$.
Then $r_\epsilon$ fulfills the above assumptions and furthermore $\exists C>0$ such that $\supp r_\epsilon \subseteq [-C,C]$ $\forall 0<\epsilon\leq 1$. Obviously $r_\epsilon\to r$ in law if $\epsilon\searrow 0$. Let $\mu_{g,\epsilon}^{\Lat}$ be the lattice measure (\ref{3.7eqa})
with $r$ replaced with $r_\epsilon$. To prove $\int_{\R^m}\tilde H\, d\mu^\Lat_{g,\epsilon}\to \int_{\R^m}\tilde H\, d\mu_g^\Lat$ as $\epsilon\searrow 0$ it suffices to 
show the related statement for the free measures with $v=0$ as the multiplication of a continuous, exponentially bounded observable $\tilde H$ with a Gibbs factor $e^{-U_g^\Lat}$ again gives an
observable of this type. As the single spin sites then are decoupled, we can identify $\Lat$ with $\{1,\ldots,m\}$ without loss of information.  Thus, by (\ref{3.11eqa}),
\begin{equation}
\label{3.13eqa}
\int_{\R^m}\tilde H d\rho_\epsilon^{\otimes m}=e^{-mz\lambda^d}\sum_{N_1,\ldots,N_m=0}^\infty {(z\lambda^d)^{N_1+\cdots+N_m}\over N_1!\cdots N_m!}\int_{\R^m}\tilde H\, d(\rho^{*N_1}_\epsilon\otimes\cdots\otimes \rho^{*N_m}_\epsilon).
\end{equation} 
The integrals on the right hand side converge to the related integrals with $\epsilon$ dropped as $\epsilon \searrow 0$: $r_\epsilon\to r$ in law implies $r^{*n}_\epsilon\to r^{*n}$ in law and the support of
$r^{*n}_\epsilon$ is contained in $[-nC,nC]$ for $1\geq\epsilon>0$. As $\tilde H$ restricted to $\times_{j=1}^m[-N_jC,N_jC]$ is bounded and continuous, the convergence of the integrals follows. 

In order to show that the r.h.s. of (\ref{3.13eqa}) converges to the left hand side of (\ref{3.12eqa}) with $\epsilon$ dropped, it only remains 
to establish a summable upper bound. We notify that $|\tilde H(s_1,\ldots,s_m)|\leq Ke^{\kappa(|s_1|+\cdots+|s_m|)}\leq Ke^{\kappa C (N_1+\cdots +N_m)}$ for $(s_1,\ldots,s_m)\in\supp \left[\otimes_{j=1}^m r_\epsilon^{*N_j}\right]$, $0<\epsilon\leq1$. 
Replacing the integrals on the r.h.s. of (\ref{3.13eqa}) with the latter $\epsilon$-independent term, we obviously get a finite sum. This finishes the proof.  
\kasten

\section{The thermodynamic limit for the grand canonical measure}

Nelson's strategy for the proof of the existence of the thermodynamic limit of $P(\phi)_2$ Euclidean quantum field theories consists of
two basic elements: Monotonicity and upper bounds for certain expectation values \cite{Si}. The FKG inequalities, established in the previous section,
give us both elements of Nelson's strategy for the class of models under consideration, as we will prove below. First we however require one further
technical assumption:
\begin{assumption}
\label{4.1assu}
The linearly bounded, concave function $v\in C^2(\R)$ defining the interaction is monotonically falling, i.e. $v'\leq 0$.
\end{assumption}

The point of Assumption \ref{4.1assu} is that it renders $-U_g$ to be monotonically increasing in the following sense: From $G\geq 0$ and (\ref{2.4eqa})
it follows that for $\eta,\gamma\in\Gamma$, $\eta\leq\gamma$ (i.e. $\eta(A)\leq \gamma(A)$ $\forall A\in{\cal B}(\R^d)$) $\Rightarrow$ $G*\eta\leq G*\gamma$ $\Rightarrow$ $-U_g(\eta)\leq -U_g(\gamma)$.

 The Assumption \ref{4.1assu} at the first look might look more restrictive than it is: Given a $v\in C^2(\R)$ linearly bounded and concave that is not monotonically falling, one can do the following replacements
\begin{eqnarray}
\label{4.1eqa}
v(\phi)&\to& v(\phi)-b\phi,~~ b\geq \|v'\|_\infty\, ,\nonumber \\
dr(s)&\to& {e^{-s b\beta \|G\|_{L^1(\R^d,dx)}}\over \int_{[-C,C]}e^{-sb \beta \|G\|_{L^1(\R^d,dx)}}\,dr(s)}\, dr(s)\, , \\
z&\to& z \int_{[-C,C]}e^{-sb\beta\|G\|_{L^1(\R^d,dx)}}\, dr(s) \,.\nonumber
\end{eqnarray}
Heuristically, the re-defined system, which now fulfills Assumption \ref{4.1assu}, and the original one describe the same physics. The only difference is that
a linear term, i.e. a self-energy term proportional to the charge $s$ of the particle, see (\ref{2.1eqa}), has been subtracted from
the interaction potential and has been added to the chemical potential.
In our case, where we deal with particles of variable charge $s\in[-C,C]$, this means that on the one hand we have to re-define the relative chemical
potentials of particles with different charges, i.e. to re-define $r$, and to re-define the over-all activity $z$ on the other hand. This explains equation (\ref{4.1eqa}).

A more technical issue however is touched by (\ref{4.1eqa}) which in a large sense can be seen as the issue of boundary conditions. To explain this, let us for
simplicity chose $r=(1/2)(\delta_{+1}+\delta_{-1})$ and that $v$ is concave with $v(-\phi)=v(\phi)$. The system then is formally invariant under the replacement $\eta\to-\eta$. The re-definitions
(\ref{4.1eqa}) break this invariance if the infra-red limit $g\nearrow\beta$ has not been taken yet: The effect of the re-definition of the energy density $v$ leads to a gain of potential 
energy of the "$+$" particles w.r.t. the "$-$" particles only in the neighborhood of $\supp g$, $g$ being the IR-cut-off function, whereas the gain in the relative chemical potential for "$-$" charged particles 
takes place everywhere. Hence, outside the support of $g$, negatively charged particles dominate. This can be interpreted as boundary  conditions of "$-$"-type. 

A related discussion with Assumption \ref{4.1assu}
formulated for $v$ monotonically increasing and $b$ in (\ref{4.1eqa}) replaced with $-b$ would then lead to "$+$"-type boundary conditions. For data leading to a phase transition (in the language of
the ferromagnetic spin systems of Section 3 this would be spontaneous magnetization) "$+$" and "$-$" type boundary conditions probably cause different phases. But a detailed discussion of this issue is beyond the scope of this article.

From now on we consider only energy densities $v$ such that Assumption \ref{4.1assu} holds. At first we want to establish monotonicity of expectation values in $g$ for monotonically increasing function:

\begin{proposition}
\label{4.1prop}
Let $F\in\hat {\cal O}(\Gamma)$ and $g_1\leq g_2$, $g_1,g_2\in C_0(\R^d)$, $g_1,g_2\geq 0$. Then,
\begin{equation}
\label{4.2eqa}
\int_\Gamma F\, d\mu_{g_1}\leq \int_\Gamma F\, d\mu_{g_2}\, .
\end{equation}
\end{proposition}
\noindent {\bf Proof.}  It suffices to show that for $F\in\hat{\cal O}(\Gamma)$, $g,f\in\C_0(\R^d)$, $g,f\geq 0$, we get
$(\partial_+\int_\Gamma F\, d\mu_{g}/\partial f)=\lim_{t\searrow 0}(\int_\Gamma F\, d\mu_{g+tf}-\int_\Gamma F\, d\mu_{g})/t\geq 0$. 

We note that $\lim_{t\searrow 0}(e^{-U_{g+tf}}-e^{-U_g})/t=-U_fe^{-U_g}$ pointwisely on $\Gamma$ and that the differential quotient
on the left hand side for $0<t\leq 1$ has a $L^p(\Gamma,\mu_0)$-upper bound, e.g. $b\langle|\eta|,f+g\rangle e^{b\langle|\eta|,f+g\rangle}$. 
Hence one gets by application of the quotient rule that
\begin{eqnarray}
\label{4.3eqa}
{\partial_+\int_\Gamma F\, d\mu_{g}\over \partial f}&=&\left. \partial_+\left[{\int_\Gamma Fe^{-U_g}\, d\mu_0\over \int_\Gamma e^{-U_g}\, d\mu_0}\right]\right/\partial f\nonumber\\
&=&\int_\Gamma F(-U_f)\, d\mu_g-\int_\Gamma F\, d \mu_g\int_\Gamma (-U_f)\, d\mu_g\, .
\end{eqnarray} 
We notify that by Assumption \ref{4.1assu} $(-U_f)$ formally is of monotonic increase, hence the right hand side of (\ref{4.3eqa}) formally is non-negative by Theorem \ref{3.1theo}, as required.

To make this argument rigorous, one could approximate $(-U_f)$ with observables in $\hat{\cal O}(\Gamma)$. Here we use the lattice approximation instead. First we recall that $(-U_f^\Lat)\in\hat {\cal O}(\R^m)$ by
Assumption \ref{4.1assu}. Hence, the right hand side of (\ref{4.3eqa}) is non-negative when going to the lattice, cf. Proposition \ref{3.4prop}. By Propositions \ref{3.1prop}--\ref{3.3prop} products of terms like
$e^{-U_{g,\Lambda}^\lambda}$, $(-U_{f,\Lambda}^\lambda)$ and $F_\Lambda^\lambda$ converge in $L^p(\Gamma,d\mu_0)$ to the corresponding products of $e^{-U_g}$, $(-U_f)$ and $F$ when first $\Lambda\nearrow\R^d$, then 
$\lambda\searrow 0$ and finally an approximation of $G$ by functions of $C_0(\R^d)$ is used. Thus, the right hand side of (\ref{4.3eqa}) can be approximated with non-negative expressions and is thus non-negative.     
\kasten

The next proposition gives upper bounds by comparison of the interacting ensemble with a gas of non-interacting particles with a different (space-dependent) activity. Let $\mu_{0,g}$ be the measure on $(\Gamma,{\cal B}(\Gamma))$ obtained
by replacing $v(\phi)$ in the definition of $\mu_g$ with the linear function $-b\phi$. 

\begin{proposition}
\label{4.2prop} Let $F\in\hat{\cal O}(\Gamma)$. Then, 
\begin{equation}
\label{4.4eqa}
\int_\Gamma F\, d\mu_g\leq \int_\Gamma F\, d\mu_{0,g}< M~~\forall g\in C_0(\R^d),0\leq g\leq\beta
\end{equation}
where $M$ is finite and depends only on $F$, $z,C,b,\|G\|_{L^1(\R^d,dx)}$ and $\beta$.
\end{proposition}
\noindent {\bf Proof.} For $\alpha\in[0,1]$ let $\mu_{\alpha,g}$ be the measure obtained by replacing $v(\phi)$ in the definition of $\mu_g$ by
$\alpha v(\phi)-(1-\alpha)b\phi$. This density is again concave for $\alpha\in[0,1]$, hence $\mu_{\alpha,g}$ fulfill the FKG correlation inequalities. 
Obviously, $\mu_{\alpha,g}$ interpolates between $\mu_{0,\alpha}$ and $\mu_{1,g}=\mu_g$.
In order to prove the first inequality in (\ref{4.4eqa}) it is sufficient to prove $d\int_\Gamma F\, d\mu_{\alpha,g}/d\alpha\leq 0$.

Let $U_{\alpha,g}$ be the potential corresponding to the energy density $\alpha v(\phi)-(1-\alpha)b\phi$ and let $U_{\alpha,g}'=dU_{\alpha,g}/d\alpha$, i.e.
$U_{\alpha,g}'$ is the potential with IR-cut-off $g$ corresponding to the energy density $v(\phi)+b\phi$. By the same reasoning as in the proof of Proposition \ref{4.1prop}
one gets
\begin{equation}
\label{4.5eqa}
{d \int_\Gamma F\, d\mu_{\alpha,g}\over d\alpha}=\int_\Gamma F (-U'_{\alpha,g})d\mu_{\alpha,g}-\int_\Gamma F\, d\mu_{\alpha,g}\int_\Gamma (-U'_{\alpha,g})\, d\mu_{\alpha,g}.
\end{equation}
We note that $U'_{\alpha,g}$ is of monotonic increase in the same sense as explained for $-U_f$ in the proof of Proposition \ref{4.1prop}. By the lattice approximation Prop. \ref{3.1prop}--\ref{3.3prop} and
the FKG inequality on the lattice Prop. \ref{3.4prop} one thus gets as in the previous proposition that the r.h.s. of (\ref{4.5eqa}) is non-positive. This proves the first inequality in (\ref{4.4eqa}).

To establish the uniform bound $M$, we note that one gets by a standard calculation on Laplace transforms (see e.g. \cite{AGY2})
\begin{equation}
\label{4.6eqa}
\int_{\Gamma}e^{\langle|\eta|,h\rangle+\langle\eta,f\rangle}\,d\mu_0(\eta)=e^{z\int_{\R^d}\int_{[-C,C]}[e^{|s|h(y)+sf(y)}-1]\, dr(s)\, dy}\,,
\end{equation} 
$f,h\in L^1(\R^d,dx)$ bounded and continuous. We set $f=bG*g$ (i.e. $-U_{0,g}(\eta)=\langle\eta,f\rangle$) and we chose $h\in C_{\rm f.f.}(\R^d)$, $h\geq 0$, such that $F(\eta)\leq K e^{\langle |\eta|,h\rangle}$. Then,
\begin{eqnarray}
\label{4.7eqa}
\int_\Gamma F\, d\mu_{0,g}&\leq& K{\int_\Gamma e^{\langle |\eta|,h\rangle+\langle\eta,bG*g\rangle}\, d\mu_0(\eta)\over \int_\Gamma e^{\langle\eta,bG*g\rangle}\, d\mu_0(\eta)}\nonumber\\
&=& K e^{z\int_{\R^d}\int_{[-C,C]}[e^{|s|h(y)+sbG*g(y)}-e^{sbG*g(y)}]\, dr(s)\, dy}
\end{eqnarray}
For $ 0<g\leq \beta$, $\|G*g\|_\infty\leq \beta \|G\|_{L^1(\R^d,dx)}$. The integrand in the exponent on the r.h.s. thus is smaller than
$Rh(y)\in L^1(\R^d,dx)$ with $R=Ce^{C\|h\|_\infty+Cb\beta\|G\|_{L^1(\R^d,dx)}}$.
\kasten

We are now in the position to formulate the main theorem of this article. The notation $g\nearrow\beta$ which is needed
to formulate the infinite volume limit has the following precise meaning: For any sequence $\{g_n\}_n\in{\N}$ of non-negative functions in $C_0(\R^d)$ such that for
any $N\in \N$ $\exists n_0\in\N$ such that $g_n(x)=\beta$ for all $x$, $|x|\leq N$, and $n\geq n_0$ the limit exists and is unique. 

\begin{theorem}
\label{4.1theo}
Let $g\nearrow \beta$. Then, 

\noindent (i) There exits a uniquely determined measure $\mu_\beta$ on $({\cal S}'(\R^d),{\cal B})$ such that
$\mu_g\to\mu_\beta$ weakly.

\noindent (ii) $\mu_\beta$ is invariant under translations. If $G$ is invariant under rotations, $\mu_\beta$ is also rotation invariant. 
\end{theorem}
\noindent {\bf Proof.} (i) We notify that by Propositions \ref{4.1prop}--\ref{4.2prop}
$\lim_{g\nearrow \beta}\int_{\Gamma} F\, d\mu_g$ exists for $F\in\hat {\cal O}(\Gamma)$. 

Let now $F\in {\cal O}(\Gamma)$ such that $F(\eta)=H(\langle\eta,h_1\rangle,\langle\eta,h_2\rangle)$, $h_1,h_2\in C_{\rm f.f.}(\R^d)$ and
$H:\R^2\to\R$ in $C^2(\R^d)$ together with first its and second order partial derivatives exponentially bounded. Then $\lim_{g\nearrow\beta}\int_\Gamma F\, d\mu_g$ still exists.
In fact, one can write $F=F^\uparrow-F^\downarrow$ with $F^{\uparrow\downarrow}\in\hat {\cal O}(\Gamma)$. As the limit exists for $F^{\uparrow\downarrow}$ separately, it also exists for $F$. 
One choice of the
$F^{\uparrow\downarrow}$ is as follows: Let $H^\pm$ be the positive/negative part of a function $H=H^+-H^-$. We set
\begin{eqnarray}
\label{4.8eqa}
H^{\uparrow\downarrow}(x,y)&=&H^\pm(0,0)+\int_0^{y}(\partial_2H(0,t))^\pm\, dt\nonumber\\
&+&\int_0^x(\partial_1H(s,0))^\pm\,ds+\int_0^y\int_0^x(\partial_1\partial_2 H(s,t))^\pm ds\,dt
\end{eqnarray}
for $x,y\in\R$, $\partial_{1/2}$ the partial derivative w.r.t. the first/second argument and $F^{\uparrow\downarrow} (\eta)=H^{\uparrow\downarrow}(\langle\eta,h_1\rangle,\langle\eta,h_2\rangle)$. 

In the next step we prove weak convergence to a measure on $({\cal S}'(\R^d),{\cal B}))$ using Minlos' theorem \cite{M}:
Let $f\in{\cal S}(\R^d)$ then $f^+$, $f^-$ are non-negative functions in $C_{\rm f.f}(\R^d)$.  Then, $F_1(\eta)=\cos(\langle\eta,f^+\rangle-\langle\eta,f^-\rangle)$ and
$F_2(\eta)=\sin (\langle\eta,f^+\rangle-\langle\eta,f^-\rangle)$ are observables of the type described in the preceding paragraph. Hence the limit of the characteristic functional
${\cal C}_g(f)=\int_{\Gamma}e^{i\langle\eta,f\rangle}d\mu_g(\eta)$ converges to ${\cal C}_\beta (f)\in\C$ for $f\in{\cal S}(\R^d)$ as $g\nearrow\beta$.

${\cal C}_\beta:{\cal S}(\R^d)\to\C$ is positive definite and normalized as the limit functional of functionals with these properties.
 It remains to prove that ${\cal C}_\beta$ is continuous w.r.t. the ${\cal S}(\R^d)$-topology. 
 For fixed cut-off $g$ one gets 
 \begin{eqnarray}
 \label{4.9eqa}
 |{\cal C}_g(f)-{\cal C}_g(h)|&\leq& \sum_{\sigma,\tau=\pm}\int_\Gamma\langle\eta,(f-h)^\sigma\rangle^\tau\, d\mu_g(\eta)\nonumber \\
 &\leq& \sum_{\sigma=\pm}\left[\int_{\Gamma}\langle\eta,(f-h)^\sigma\rangle^-\,d\mu_0(\eta) +\int_{\Gamma}\langle\eta,(f-h)^\sigma\rangle^+\,d\mu_{0,g}(\eta)\right].\nonumber \\
 \end{eqnarray} 
In the last step we applied Proposition \ref{4.1prop} and \ref{4.2prop} to the decreasing observables $\langle\eta,(f-h)^\pm\rangle^-$ and the increasing
observables $\langle\eta,(f-h)^\pm\rangle^+$. As $h\to f$ in ${\cal S}(\R^d)$ it follows $(f-h)^\pm\to 0$ in $C_{\rm f.f}(\R^d)$ and the
$d\mu_0$-integrals vanish under this limit, as can be seen by dominated convergence in $L^p(\Gamma,\mu_0)$. 

We have to give a $g$-independent estimate for the vanishing of the $\mu_{0,g}$-integrals for $0\leq g\leq\beta$. One can proceed as follows:
\begin{eqnarray}
\label{4.10eqa}
\int_{\Gamma}\langle\eta,(f-h)^\pm\rangle^+\,d\mu_{0,g}(\eta)&\leq&\int_\Gamma (e^{\langle|\eta|,(f-h)^\pm\rangle}-1)\, d\mu_{0,g}(\eta)\nonumber\\
&\leq& e^{zR\int_{\R^d}(f-h)^\pm \, dy}-1\to 0
\end{eqnarray}   
as $(h-f)^+\to 0$ in $C_{\rm f.f}(\R^d)$ with $R=R((h-f)^+)$ as in the proof of Proposition \ref{4.2prop} independent of $g$ and bounded for $(h-f)^\pm\to 0$ in $C_{\rm f.f}(\R^d)$. 

We have proven that ${\cal C}_\beta$ is a characteristic functional  and $\mu_\beta$ can now be defined as the unique measure on $({\cal S}'(\R^d),{\cal B})$ with
the given characteristic functional, \cite{M}. As weak convergence is equivalent with the convergence of characteristic functionals by L\'evy's theorem, the assertion of (i) follows. 

(ii) Let $G$ be invariant under rotations and reflections. For $\eta\in\Gamma$, $\eta=\sum_{y\in{\rm supp}\eta}s_y\delta_y$, $x\in\R^d$ and $D$ an element of the orthogonal group on $\R^d$ let $\eta_{\{x,D\}}=\sum_{y\in{\rm supp}\eta}s_y\delta_{Dy+x}$ and $h_{x,D}(y)=h(D^{-1}(y-x))$, $h:\R^d\to\R$. For $F\in L^1(\Gamma,\mu_0)$
let $T_{\{x,D\}}F(\eta)=F(\eta_{\{x,D\}})$.  $\mu_0$ is invariant in distribution under Euclidean transformations, i.e. $T^*_{\{x,D\}}\mu_0=\mu_0$ with $T^*_{\{x,D\}}$ the dual action of $T_{\{x,D\}}$ (this e.g. can be deduced from the invariance of (\ref{2.2eqa}) under such transformations and
the uniqueness statement of Minlos' theorem). For the potential energy we get $U_g(\eta_{\{x,D\}})=\int_{\R^d}v(G*\eta_{\{x,D\}})\, gdx =\int_{\R^d}v((G*\eta)_{\{x,D\}})\, gdx=\int_{\R^d}v(G*\eta)\, g_{\{x,D\}^{-1}}\, dx$. Thus, $T^*_{\{x,D\}}\mu_g=\mu_{g_{\{x,D\}}}$. The assertion now follows from the equivalence of the limit $g\nearrow \beta$ and $g_{{\{x,D\}}}\nearrow\beta$. 
If $G$ is not invariant under rotations, the argument still holds for $D=1$. 
\kasten 
 
 \
 
\small
\noindent {\bf Acknowledgments.} 
Helpful discussions with Sergio Albeverio and Minoru W. Yoshida are gratefully acknowledged. I have to thank Tobias Kuna for pointing out to me Ref. \cite{GK}.
This work has been made possible through financial support of D.F.G through Project "Stochastic analysis and systems of infinitely many degrees of freedom", "Stochastic Methods in QFT" and SFB 611 A4.

\end{document}